\documentclass[aps,prl,twocolumn,superscriptaddress,preprintnumbers,nofootinbib]{revtex4-1}

\usepackage{mathrsfs} 
\usepackage{amsfonts}
\usepackage{amsmath}
\usepackage{amsthm}
\usepackage{amssymb}
\usepackage{array}
\usepackage{dcolumn}
\usepackage[normalem]{ulem}
\usepackage[svgnames]{xcolor}
\usepackage[hidelinks]{hyperref}
\hypersetup{
    colorlinks=true,
    linkcolor=NavyBlue,
    filecolor=NavyBlue,
    urlcolor=NavyBlue,
    citecolor=NavyBlue,
    }

\usepackage{graphics}
\usepackage{graphicx}
\usepackage[caption=false]{subfig}
\usepackage{adjustbox}
\usepackage{breqn}
\usepackage{fontawesome} 
\usepackage{slashed}

\def\be{\begin{equation}}
\def\ee{\end{equation}}
\def\bea{\begin{eqnarray}}
\def\eea{\end{eqnarray}}

\newcount\Comments  
\newcommand{\kibitz}[2]{\ifnum\Comments=1\textcolor{#1}{#2}\fi}

\makeatletter
\let\cat@comma@active\@empty
\makeatother

\begin{document}

\title{Millicharged Particle Production in Pulsars via the Schwinger Effect}
\author{Chris Kouvaris}
\email{kouvaris@mail.ntua.gr}
\affiliation{Physics Division, National Technical University of Athens, 15780 Zografou Campus, Athens,
Greece}

\author{Ian M.~Shoemaker}
\email{shoemaker@vt.edu}
\affiliation{Center for Neutrino Physics, Department of Physics, Virginia Tech, Blacksburg, VA 24061, USA}


\begin{abstract}

Low mass particles with small electric charges can be produced abundantly in large electric fields via the Schwinger effect. We study the production rate of such particles inside the polar gap of nearby pulsars. After production they are accelerated above MeV energies by the local electric fields. These pulsar-produced millicharged particles can be detected at Earth in low-threshold dark matter direct detection experiments. We find that the current XENONnT data constrains millicharged particles produced in the Crab pulsar to have charges less than $\mathcal{O}(10^{-6})$ for sub-eV masses.

\end{abstract}

\maketitle

\textbf{\textit{Introduction-}} Millicharged particles (MCPs) are a natural candidate for physics beyond the Standard Model~\cite{Davidson:2000hf}. They arise in many dark sector scenarios in which either a fermion or scalar is coupled to a new $U(1)$ vector field which kinetically mixes with the photon, $\epsilon F'_{\mu\nu}F^{\mu \nu}$~\cite{Holdom:1985ag}. In the limit that this new vector is massless, the new fermion or scalar will have an ordinary electric charge with a value suppressed by $\epsilon$. In recent years, a variety of new search strategies have been invoked to study MCPs across many orders of magnitude in MCP mass utilizing both terrestrial~\cite{Ahlers:2007qf,DellaValle:2014xoa,Magill:2018tbb,Berlin:2018bsc,Kelly:2018brz,Foroughi-Abari:2020qar,Oscura:2023qch,SENSEI:2023gie,Tsai:2024wdh,Essig:2024dpa,Berlin:2024dwg,Gao:2025ykc}, astrophysical~\cite{PhysRevD.49.2114,Chang:2018rso,Harnik:2020ugb,Plestid:2020kdm,ArguellesDelgado:2021lek,Berlin:2021kcm,Li:2024pcp,Wu:2024iqm}, and cosmological sources~\cite{Davidson:2000hf,Gan:2023jbs,Iles:2024zka,Berlin:2025hjs,Berlin:2025btf}. 

In this paper we will explore the consequences of millicharged particles on pulsars, in particular their production via the Schwinger effect. After production they will be accelerated to $\mathcal{O}({\rm MeV})$ energies, making them detectable at Earth. Previous work examining the impact of MCPs on pulsars include the time delay of electromagnetic signals induced by background millicharged particles~\cite{Caputo:2019tms}.

Due to Quantum Mechanics, the vacuum is a dynamic environment with particle-antiparticle pairs coming and going. However in the presence of a strong electromagnetic field, these virtual particles can go on-shell. If the potential energy of a virtual electron, $qEd$, exceeds the rest mass of the electron on Compton wavelength scales $d\sim m^{-1}$, electron-positron pairs can be copiously produced. Thus there is a critical field value for spontaneous pair production from the electromagnetic vacuum
\be
q E/m \gtrsim m \Longrightarrow E\gtrsim E_{{\rm crit}} \equiv  m^{2}/q
\label{eq:basic}
\ee
This process is dubbed the Schwinger effect~\cite{PhysRev.82.664}, and from Eq.~(\ref{eq:basic}) it is apparent that hypothetical particles with small charges can be produced with smaller electric fields~\cite{Berlin:2020pey}. Note that it has been previously demonstrated that the Schwinger mechanism within standard QED does not appreciably alter pulsar dynamics~\cite{PhysRevD.14.340}.

\textbf{\textit{Schwinger Production-}} In what follows we will consider new fermions with a charge $q_{F} = e\epsilon$, where $e$ is the electron charge and $\epsilon$ is a small, dimensionless number~\footnote{Note that when $B/E \gg1$ pair production rates for charged scalars is suppressed by a factor $e^{-\pi B/E}$.}. In a region where $\bf{E}$ and ${\bf B}$ are parallel (and both $E \cdot B \neq0, (E^{2} - B^{2}) \neq 0$), the imaginary part of the one loop effective Lagrangian is~\cite{PhysRev.82.664,Kim:2003qp}
\be 
2 {\rm Im}(\mathcal{L})= \frac{q_{F}^{2} EB}{(2\pi)^{2}}~\sum_{n=1}^{\infty}\frac{1}{n}\coth \left(\frac{n \pi B}{E} \right) \exp\left(-\frac{n \pi m^{2}}{q_{F}E} \right) 
\label{eq:lagr}
\ee
The fermion pair production rate per unit volume has been shown to be the first term~
in this series~\cite{Nikishov:1970br,Cohen:2008wz}
\be
\Gamma_{F}= \frac{q_{F}^{2} EB}{(2\pi)^{2}}\coth \left(\frac{ \pi B}{E} \right) \exp\left(-\frac{\pi m^{2}}{q_{F}E} \right) 
\label{eq:rate}
\ee


In the pioneering work of Goldreich and Julian~\cite{Goldreich:1969sb} it was demonstrated that a rotating neutron star with a magnetic field must be surrounded by a magnetosphere. Considering the neutron star as a perfect conductor, an electric field is induced on the star in order to have net zero Lorentz force on a charged particle inside the star in the co-rotating frame. This electric field extracts electrons and protons from the surface of the star, accelerating them to high velocities. The amount of charge flowing out from the star is so-called Goldreich-Julian current.  However as it was shown by Ruderman and Sutherland~\cite{Ruderman:1975ju}, protons might not be so easy to extract from the star in contrast to what happens for electrons. Consequently the obstruction of the positive current could lead to the formation of ``polar gaps'' close to the star's surface. Within the gap region there is strong electric and magnetic field and practically very few particles. Given a strong enough electric field and once the height of the gap has grown sufficiently, the mean free path of curvature radiation photons to break in positron-electron pairs becomes comparable to the gap's height. As a result a bunch of electron-positron pairs accelerate within the gap. In turn they produce more curvature photons which lead to more pair production. This leads temporarily to the discharge of the gap. Once this happens the electric field drops to zero and the pair production stops. The gap grows again with an increasing electric field and height until it reaches the threshold once again for pair production. This repeated cycle is eventually responsible for the pulsar emission.

This gap volume is a cylindrical region of (maximum) height $h$ estimated to be~\cite{Ruderman:1975ju}
\be 
h \simeq 5 \times 10^{3}~{\rm cm}~\rho_{6}^{2/7}P^{3/7}B_{12}^{-4/7},
\label{eq:height}
\ee
where $\rho_{6} \equiv \rho/(10^{6}~{\rm cm})$ is the radius of curvature of an electron trajectory, $P$ is the pulsar period in seconds, and $B_{12} \equiv B/(10^{12}~{\rm Gauss})$ is the radial component of the surface field strength.  
Ruderman and Sutherland~\cite{Ruderman:1975ju} found that the potential difference across the gap is 

\be
\Delta V  \simeq 1.6\times 10^{12}{\rm V}~B_{12}^{-1/7}P^{-1/7}\rho_{6}^{4/7}
\ee

The volume of the polar gap can be estimated as
\be
V = \pi \left(r_{p}^{2}-r_{c}^{2}\right)h
\label{eq:vol}
\ee
where $r_{p}=R\sqrt{\omega R} \simeq 10^{4}~{\rm cm}~P^{-1/2}$ and $r_{c}=(2/3)^{3/4}r_p \simeq 1.4 \times 10^{4} ~{\rm cm}~P^{-1/2}$~\cite{Ruderman:1975ju}. $r_p$ is the distance from the pole where the last magnetic field lines that barely cross the light cylinder start from, while $r_c$ represents the distance from the pole where the critical field lines that cross the light cylinder at right angle start from. Here we assume that the magnetic and rotation axes are parallel. In this case electrons are emitted from the region around the pole up to radius $r_c$ while the gap is formed in the annulus between $r_c$ and $r_p$. If for example the magnetic and rotation axes are antiparallel, then the picture is reversed and the gap forms with the region of $r_c$. In that case the volume will be given by $\pi r_c^2 h$. There is no appreciable difference in the volume and production rate of particles between the two cases. In reality, pulsars usually have  magnetic and rotation axes that are neither parallel nor antiparallel. For example the Crab pulsar that we will use to set constraints has a angle around $45^\circ$ which is none of the above cases. Although it is hard to estimate the exact volume in this geometry, it is expected that will fall somewhere between the two extreme aforementioned cases and therefore, it will not affect our already conservative estimate to assume a parallel configuration.

We estimate the electric field in the gap to be 
\be
E \simeq 2 \frac{\Delta V}{h} \simeq \left( 6.4 \times 10^{8}~{\rm V/cm}\right) ~B_{12}^{3/7}P^{-4/7}\rho_{6}^{2/7}
\label{eq:electric}
\ee
As a result, after traversing a distance $h$ one can expect a particle with charge $e\epsilon$ to acquire an energy
\be
\mathcal{E}_{X} \simeq 1.6 \times 10^{5}~{\rm eV}~\left(\frac{\epsilon}{10^{-7}}\right)~B_{12}^{-1/7}P^{-1/7}\rho_{6}^{4/7}
\label{eq:Emax}
\ee
Note that Eq.(\ref{eq:Emax}) implies that reaching the smallest values of $\epsilon$ requires experimental sensitivity to the low energies.   

\begin{figure}[t!]
    \includegraphics[width = 0.49 \textwidth]{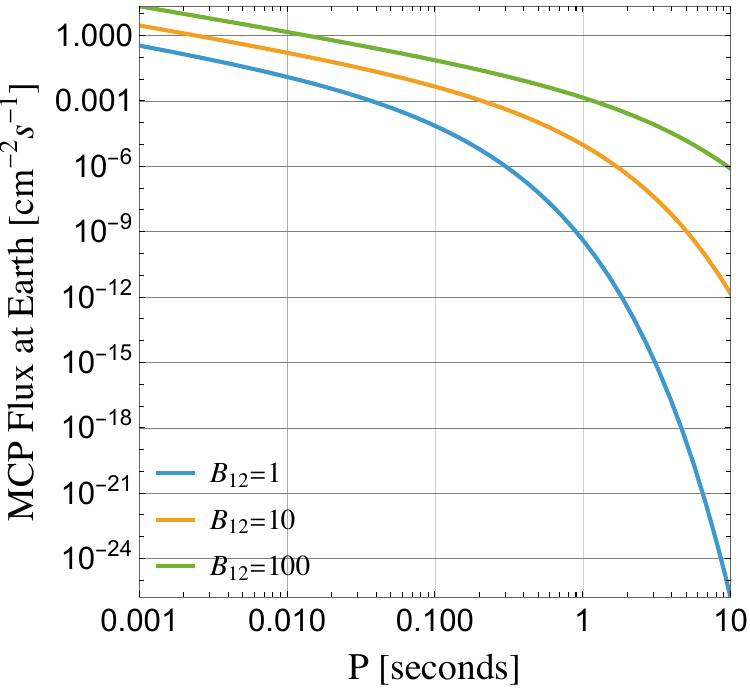}
    \caption{Here we display the MCP flux arriving at Earth from a pulsar at a distance of $L=1$ kpc. In this representative example we have fixed $\epsilon =10^{-7}$ and $m_{X} = 0.1$ eV.  }
    \label{fig:flux}
\end{figure}


Notice that as a result of the exponential in Eq.~(\ref{eq:rate}), the production of MCP with masses greater than a critical value, $m_{{\rm crit}}$, is strongly suppressed. By inspection this critical mass is roughly
\be
m_{{\rm crit}} \equiv \sqrt{\frac{q_{F} E}{\pi}}
\label{eq:mcrit}
\ee
Now that we know the approximate electric field strength we can estimate the critical MCP mass scale, above which we expect the production to become very suppressed. From Eq.~(\ref{eq:mcrit}), we find
\be
m_{{\rm crit}} \simeq 0.1~{\rm eV}~\sqrt{\frac{\epsilon}{10^{-7}}}
\ee
Thus assuming such states are produced isotropically in a volume $V$ of a pulsar gap at a distance $L$ from Earth, we expect a flux at Earth
\be
\Phi_{F} = \frac{V\Gamma_{F} }{4\pi L^{2} },
\label{eq:Phi}
\ee
where the volume $V$ is given by Eq.~(\ref{eq:vol}), and the rate is given by Eq.~(\ref{eq:rate}) with the electric field from Eq.~(\ref{eq:electric}). We plot Eq.~(\ref{eq:Phi}) in Fig.~\ref{fig:flux} as a function of pulsar period for a variety of pulsar magnetic field values, where we notice that the largest fluxes occur for the smallest periods and largest magnetic fields.  
%


%
We should stress here that the assumption of isotropy in the emission of the particles gives a conservative estimate of the actual flux of MCPs arriving on Earth. In principle MCPs are emitted within a cone of the angle opening $\theta_p$ of the polar gap given by $\sin \theta_p = r_{p}/R$. This will enhance the flux of MCP within the cone instead of the isotropic distribution  by a factor $\sim 4/\theta_p^2$ (for small values of $\theta_p$). However, one should take into account the fact that the cone of emission is rotating and the flux on Earth is active only during the passing of the cone from the line of sight. This reduces the enhancement in the flux. An estimate of the duty cycle results to a final enhancement of $\sim 2/\theta_p$ with respect to the isotropic emission. The flux enhancement ranges from a factor of  $\sim 4$ for millisecond pulsars to a factor of $\sim 100$ for a pulsar with a period of 1 second. Since fast rotating pulsars give the most stringent constraints and for reasons of simplicity, we assume isotropic emission, understanding that this gives a true conservative and model independent lower limit for the MCP flux.

In principle the MCP spectrum is affected by the continuous
charging/discharging of the polar gap, a process that according to Ref.~\cite{Ruderman:1975ju} takes of the order of microseconds to complete.
Since the detailed variation of the electric field depends on the
assumed polar gap model, we leave for a future work a more elaborate
study of this effect. Here we consider a time averaged electric field
over the cycle introducing only a minor error in our estimate of the MCP
flux.

\textbf{\textit{The Crab Pulsar and Other Pulsar Sources-}} We may now compute expected fluxes of MCPs based on inputting specific pulsar parameters into the above formalism. As a representative example, we will take Crab pulsar (PSR J0534+2200), whose properties have been well measured over many years. The timing of the beamed radio signal yields a very precisely measured period~\cite{Manchester:2004bp,Lyne:1993xah}
\be 
P_{{\rm Crab}} = 0.0338238880741~{\rm s}
\ee
Although less precisely known since it is derived from other measured parameters the Crab pulsar surface magnetic field is bounded to be~\cite{Philippov:2013aha}
\be
B_{{\rm Crab}} > 8.5 \times 10^{12}~{\rm Gauss}
\ee
And lastly the distance from Earth is also well-known~\cite{Lin:2023erb}
\be
L_{{\rm Crab}} = 1.9^{+0.22}_{-0.18}~{\rm kpc}
\ee
With the above values we find that the Crab pulsar produces the following fiducial MCP flux at Earth and maximum MCP energy (assuming $\epsilon = 10^{-6}$ and $m_{X} = 0.1$ eV)~\footnote{We note that the energy flux (or intensity) $\Phi \times \mathcal{E}_{X}$ can be expressed in so-called ``Crab'' units which are a conventional astrophotometrical unit, where $1~{\rm Crab} \equiv 2.4 \times 10^{-8}~{\rm cm}^{-2}~{\rm s}^{-1}$. In the above example, the MCP intensity at earth is $(\Phi \times \mathcal{E}_{X}) \simeq 2000~{\rm Crab}$. }
\bea 
\Phi_{{\rm Crab}} &\simeq& 1.3~{\rm cm}^{-2}~{\rm s}^{-1} \\
\mathcal{E}_{X} &\simeq& 24~{\rm MeV}
\eea

We have checked that of the pulsars in the ATNF catalog~\cite{Manchester:2004bp} with known magnetic fields, the Crab Pulsar produces the largest MCP flux by around a factor of 100 times larger than the next brightest pulsar. We note that after the Crab pulsar the next two brightest are PSR J1513-5908 and PSR J1846-0258 which are nearly degenerate in MCP flux.


\textbf{\textit{Magnetic Field Impact}} Next we will check if the galactic magnetic field can play a role in MCP propagation. The relativistic Larmor radius is
\be
r_{L} = \frac{\gamma mv_{\perp}}{qB} = \frac{\mathcal{E}_{X}(v_{\perp}/c)}{e \epsilon B_{Gal} } \simeq 4 \times 10^{-6}{\rm kpc}~\left(\frac{1\mu G}{B_{Gal}}\right)\left(\frac{v_{\perp}}{c} \right)
\ee
Thus for a ``pure'' MCP with direct photon couplings, this will have an important impact on MCP trajectories and the expected flux at Earth. However ``effective MCPs" in a dark photon scenario, will appear equivalent to a MCP on distances much smaller than $m_{A'}^{-1}$. Thus for $m_{A'}^{-1} < r_{L}$, the Larmor effects will be absent. For the above parameters this happens when the dark photon obeys, $m_{A'} \gtrsim 10^{-21}$eV. 

At the same time however, for the dark photon to be essentially massless on the length scales relevant for the pulsar, it should have a range at least as large as the gap height in Eq.~\ref{eq:height}. As a result the dark photon mass should be $m_{A'} \lesssim  10^{-9}$ eV in order to behave as an electromagnetically coupled state on pulsar scales. Finally note that for dark photon masses below roughly $m_{A'} < 10^{-14}$ eV the kinetic mixing parameter constraints are weak enough to induce an effectively millicharged particle with $\mathcal{O}(\epsilon) \sim 10^{-6}$ on pulsar length scales~\cite{Caputo:2021eaa}.


\textbf{\textit{Direct Detection Constraints on the MCP Flux from the Crab Pulsar-}} Given the relatively low energy of the MCP flux produced in pulsars, electron scattering in direct detection experiments will offer the best sensitivity. In this section we will derive bounds on the MCP flux based on the electron recoil data collected by XENONnT~\cite{XENON:2022ltv} based on 1.16 ton-yr of exposure. The XENONnT collaboration found no excess of events in their analysis of electron recoil events between 1 and 30 keV, and achieved significantly smaller background rates compared to previous experiments. We note that similar bounds can be derived from the PandaX-4T data~\cite{PandaX:2024cic} though the background rate is slightly larger for PandaX-4T. In the future it could be interesting to look at the ionization-only data from XENONnT~\cite{XENON:2024znc}.

\begin{figure}[t!]
    \includegraphics[width = 0.45 \textwidth]{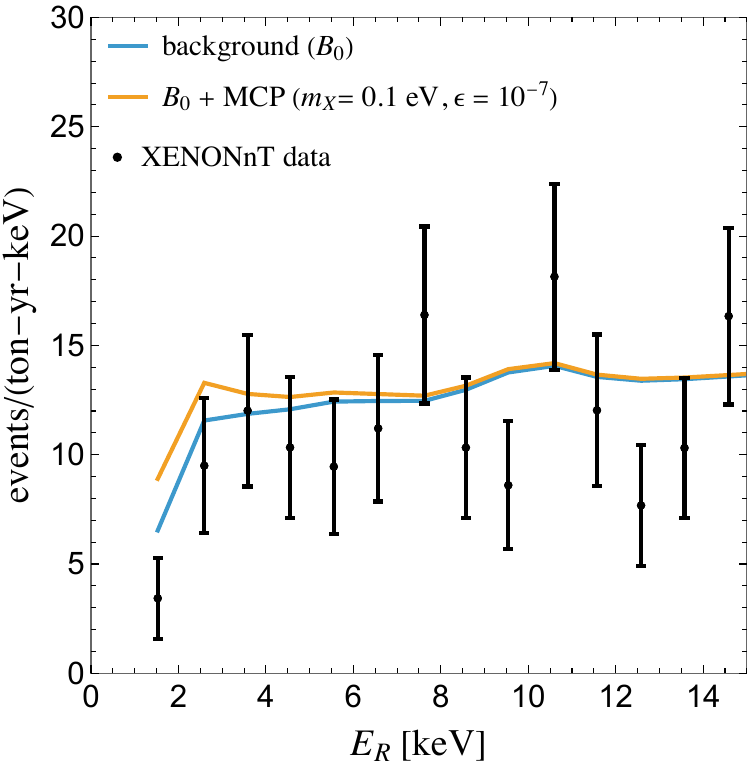}
   \includegraphics[width = 0.48 \textwidth]{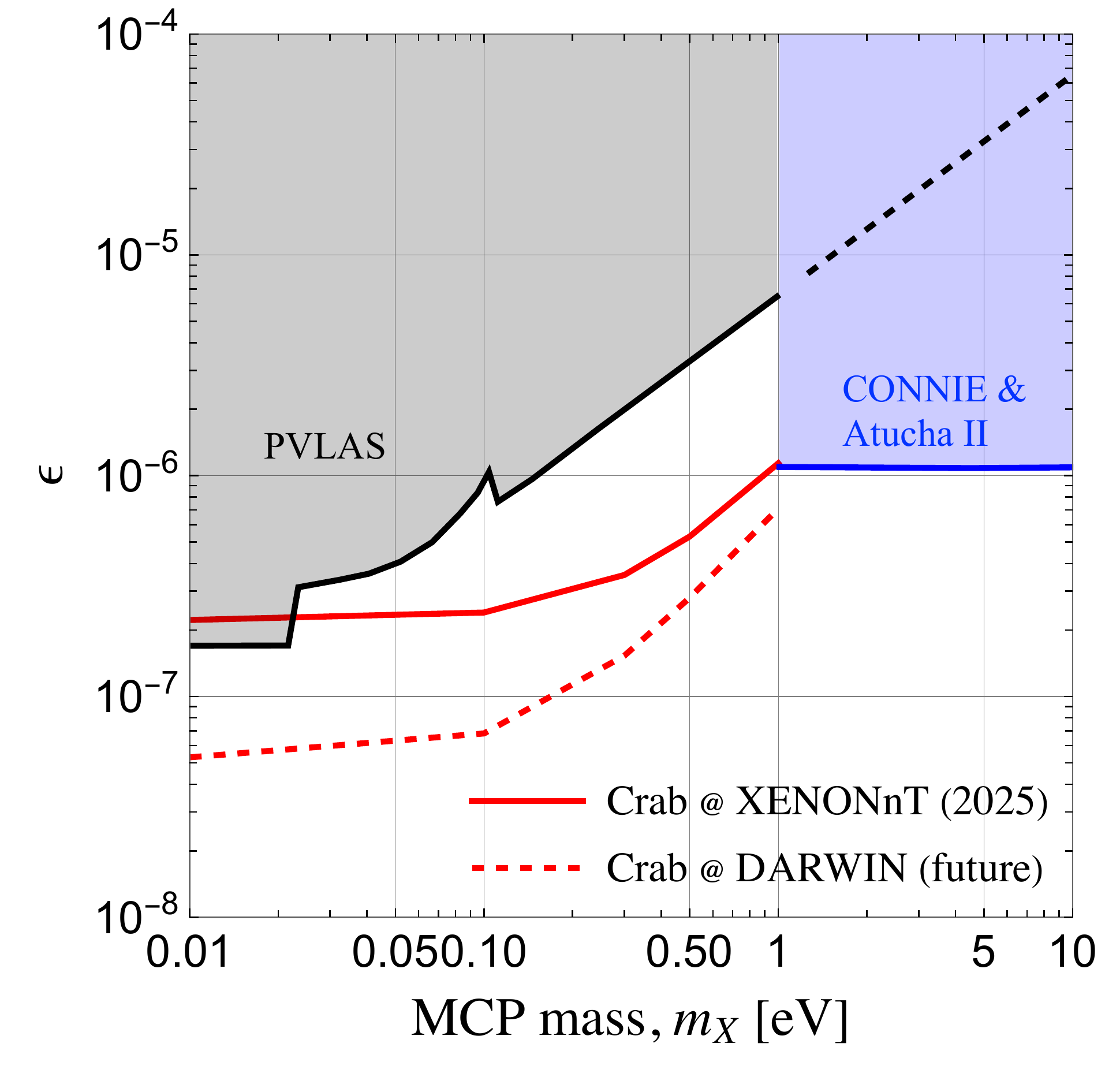}
    \caption{ ({\it Upper panel}:) An example XENONnT electron recoil event rate from MCPs produced in the Crab Pulsar. ({\it Lower panel:}) The corresponding 90$\%$ CL bounds on the MCP properties with recent XENONnT data~\cite{XENON:2022ltv}, and a projection for DARWIN. Existing bound from the PVLAS collaboration are shown in gray, while the dashed curve is a naive extrapolation of their published bounds~\cite{Ahlers:2007qf,DellaValle:2014xoa}, along with the recent bounds from CONNIE and Atucha II~\cite{CONNIE:2024off}.} 
    \label{fig:bounds}
\end{figure}

To make estimates of the direct detection bounds, we use the electron elastic scattering cross ~\cite{Essig:2024dpa} (assuming $E_{X} \gg E_{e},m_{X},m_{e}$):
\be
\frac{d\sigma}{dE_{e}} = \pi \alpha^{2} \epsilon^{2}~\frac{2 E_{X}^{2}m_{e}+ E_{e}^{2}m_{e} - E_{e}(m_{X}^{2}+ m_{e}(2E_{X} + m_{e}))}{E_{e}^{2}(E_{X}^{2} -m_{X}^{2})m_{e}^{2}}
\ee

In order to compute the event spectrum we fold in the above cross section and flux
\be
\frac{dR}{dE_{e}} =Z_{{\rm eff}}(E_{e})~ {\rm Eff}(E_{e}) \int_{E_{min}(E_{e})}dE_{X} \frac{d \sigma}{dE_{e}} \frac{ d \phi}{dE_{X}} ,
\ee
where $ {\rm Eff}(E_{e})$ is the experimental efficiency function given in Fig.~1 of Ref.~\cite{XENON:2022ltv}, and $Z_{{\rm eff}}(E_{e})$ is the number of target electrons which can be ionized for a given $E_{e}$~\cite{thompson_x-ray_nodate}.    

%
%


Now that we have the event rate, we can add this to the background from XENONnT~\cite{XENON:2022ltv} and compare with our signal (accounting for detector efficiency). This comparison is shown in the top panel of Fig.~\ref{fig:bounds} where we see by eye that current XENONnT data offers sensitivity at the $\mathcal{O}(\epsilon) \sim 10^{-7}$ level. This is confirmed by the log-likelihood analysis we perform in which we compare the observed data with the Crab pulsar flux of MCPs, with the resulting constraints shown in the bottom panel of Fig.~\ref{fig:bounds} where we observe that even with present data the non-observation of MCPs from the Crab pulsar in XENONnT data provides new constraints on MCPs.   



\textbf{\textit{Suppressed Astrophysical and Cosmological bounds-}}Strong bounds from astrophysics can constrain some millicharged particle scenarios. These include stellar cooling~\cite{Vinyoles:2015khy}, red giant stars~\cite{Dobroliubov:1989mr,Davidson:2000hf}, SN1987A~\cite{Mohapatra:1990vq}, horizontal branch stars, and white dwarfs. In addition, there are also very strong cosmological bounds which apply to scenarios in which the MCP has a direct EM charge~\cite{Gan:2023jbs}. Furthermore there are strong bounds on relic MCPs which can constitute all or a fraction of the dark matter~\cite{Dunsky:2018mqs}.

However, they rely on specific model assumptions which not all scenarios obey. In particular there are at least two classes of models which successfully evade these bounds: In the first class of models a scalar $\phi$ couples both to the MCP with a term $g_{\chi} \bar{\chi}\chi$ (where $\chi$ is the MCP and $g_{\chi}$ a coupling) and to baryons with a similar term $g_N \bar{N}N$. In the interior of stars where the baryon density is sufficiently large, $\phi$ gets an expectation value $\langle \phi \rangle = g_N n_B/m_{\phi}^2$ where $n_B$ and $m_{\phi}$ are the baryon density and mass of $\phi$ respectively. This expectation value acts as a mass source term for the MCP changing  the MCP mass from $m_{\chi}$ to $m_{\chi}+ g_{\chi}\langle \phi \rangle$. For appropriate parameters the effective mass of the MCP can become so large that the MCPs might not be produced thermally in stars, thus evading all constraints based on fast cooling due to thermal light particle emission from the stars
~\cite{DeRocco:2020xdt}.  A second class of models involves a dark photon scenario in which the dark photon acquires a large effective mass in plasmas~\cite{Berlin:2020pey}. Our present scenario utilizes MCP production in the vacuum gap of pulsars, which should not suffer from either potential suppression mechanism. This highlights the importance and complementarity of laboratory searches for MCPs.






\textbf{\textit{Energy Losses-}} It has been previously argued that MCPs will dominantly lose energy via ionization~\cite{Dunsky:2018mqs}, where the stopping power is 
\be
\frac{dE}{dx} \simeq 0.15~{\rm MeV}~{\rm cm^{2}/g}~\left(\frac{e \epsilon}{\beta}\right)^{2}\left(\frac{Z/A}{1/2}\right)~\ln{\left(\frac{2m_{e}\gamma^{2}\beta^{2}}{10 Z~{\rm eV}}\right)}
\ee
Taking $\epsilon = 10^{-6}$, $m_{X} = 0.1$ eV, and $E_{X} = 10~{\rm MeV}$ we find that the stopping effects amount to an energy loss of $\sim 10^{-7}~{\rm MeV}$ through a kilometer of rock. In the region of parameter space relevant to this work then, the MCPs can be taken to be effectively transparent to ordinary matter. We note that Ref.~\cite{Du:2022hms} also argues that energy losses are small in this regime (around $10^{-8}$ MeV/km in rock). Using the scalings with $\epsilon$ and the MCP mass in Ref.~\cite{Hu:2016xas} we have also confirmed that bremstrahlung energy losses are also negligible for our charges and masses. 

We can also verify that stopping effects from the magnetosphere are insignificant by estimating the total change in MCP energy as it traverses the magnetosphere. This can be estimated via $\Delta E = \int (dE/dx)\rho(r)dr$, where the Goldreich-Julian density~\cite{Goldreich:1969sb} is $\rho(r) = 7 \times 10^{-2}~{\rm cm}^{-3}~m_{e} (B/P)$, where $B$ is in Gauss and $P$ is in seconds. For the same MCP example used above ($\epsilon = 10^{-6}$, $m_{X} = 0.1$ eV, and $E_{X} = 10$ MeV) we find that the change in MCP energy is $\Delta E \simeq 10^{-20}$ MeV.

\textbf{\textit{Conclusions-}} Here we have derived a new set of novel constraints on millicharged particles under certain assumptions (i.e. there is a polar gap in pulsars) based on unobserved events in dark matter direct search experiments that would have been produced by incoming fluxes of millicharged particles produced in the polar gap of the Crab pulsar via the Schwinger effect.  While we have focused on the Crab pulsar as a source, nearby pulsars with large magnetic fields and short periods may provide even better sensitivity to MCPs. Barring that, future large detector with low-thresholds and low-background rates can improve the constraints on MCPs from the Crab pulsar.    

\section*{Acknowlegements}

We are very grateful to Don N. Page for clarifying correspondence on the Schwinger Effect. IMS is supported by the U.S. Department of Energy under the award number DE-SC0020262.

\bibliographystyle{apsrev4-1.bst}
\bibliography{bibliography.bib}

@article{PhysRevD.14.340,
  title = {Theory of pair production in strong electric and magnetic fields and its applicability to pulsars},
  author = {Daugherty, J. K. and Lerche, I.},
  journal = {Phys. Rev. D},
  volume = {14},
  issue = {2},
  pages = {340--355},
  numpages = {0},
  year = {1976},
  month = {Jul},
  publisher = {American Physical Society},
  doi = {10.1103/PhysRevD.14.340},
  url = {https://link.aps.org/doi/10.1103/PhysRevD.14.340}
}

@article{PhysRev.82.664,
  title = {On Gauge Invariance and Vacuum Polarization},
  author = {Schwinger, Julian},
  journal = {Phys. Rev.},
  volume = {82},
  issue = {5},
  pages = {664--679},
  numpages = {0},
  year = {1951},
  month = {Jun},
  publisher = {American Physical Society},
  doi = {10.1103/PhysRev.82.664},
  url = {https://link.aps.org/doi/10.1103/PhysRev.82.664}
}

@article{PandaX:2024cic,
    author = "Zeng, Xinning and others",
    collaboration = "PandaX",
    title = "{Exploring New Physics with PandaX-4T Low Energy Electronic Recoil Data}",
    eprint = "2408.07641",
    archivePrefix = "arXiv",
    primaryClass = "hep-ex",
    doi = "10.1103/PhysRevLett.134.041001",
    journal = "Phys. Rev. Lett.",
    volume = "134",
    number = "4",
    pages = "041001",
    year = "2025"
}

@article{XENON:2024znc,
    author = "Aprile, E. and others",
    collaboration = "XENON",
    title = "{Search for Light Dark Matter in Low-Energy Ionization Signals from XENONnT}",
    eprint = "2411.15289",
    archivePrefix = "arXiv",
    primaryClass = "hep-ex",
    doi = "10.1103/PhysRevLett.134.161004",
    journal = "Phys. Rev. Lett.",
    volume = "134",
    number = "16",
    pages = "161004",
    year = "2025"
}

@article{Goldreich:1969sb,
    author = "Goldreich, Peter and Julian, William H.",
    title = "{Pulsar electrodynamics}",
    doi = "10.1086/150119",
    journal = "Astrophys. J.",
    volume = "157",
    pages = "869",
    year = "1969"
}

@article{Berlin:2025hjs,
    author = "Berlin, Asher and Bogorad, Zachary and Graham, Peter W. and Ramani, Harikrishnan",
    title = "{Cavendish Tests of Millicharged Particles}",
    eprint = "2510.25825",
    archivePrefix = "arXiv",
    primaryClass = "hep-ph",
    reportNumber = "FERMILAB-PUB-25-0623-SQMS-T",
    month = "10",
    year = "2025"
}

@article{Berlin:2024dwg,
    author = "Berlin, Asher and Harnik, Roni and Li, Ying-Ying and Xu, Bin",
    title = "{Millicharged Condensates on Earth}",
    eprint = "2404.16094",
    archivePrefix = "arXiv",
    primaryClass = "hep-ph",
    reportNumber = "FERMILAB-PUB-24-0002-SQMS-T, USTC-ICTS/PCFT-24-08",
    month = "4",
    year = "2024"
}

@article{Berlin:2025btf,
    author = "Berlin, Asher and Bogorad, Zachary and Graham, Peter W. and Ramani, Harikrishnan",
    title = "{Electric Accumulation of Millicharged Particles}",
    eprint = "2510.25834",
    archivePrefix = "arXiv",
    primaryClass = "hep-ph",
    reportNumber = "FERMILAB-PUB-25-0622-SQMS-T",
    month = "10",
    year = "2025"
}

@article{Essig:2024dpa,
    author = "Essig, Rouven and Li, Peiran and Liu, Zhen and McDuffie, Megan and Plestid, Ryan and Xu, Hailin",
    title = "{Probing millicharged particles at an electron beam dump with ultralow-threshold sensors}",
    eprint = "2412.09652",
    archivePrefix = "arXiv",
    primaryClass = "hep-ph",
    doi = "10.1007/JHEP04(2025)057",
    journal = "JHEP",
    volume = "04",
    pages = "057",
    year = "2025"
}

@article{Dobroliubov:1989mr,
    author = "Dobroliubov, M. I. and Ignatiev, A. Yu.",
    title = "{MILLICHARGED PARTICLES}",
    reportNumber = "RIFP-837",
    doi = "10.1103/PhysRevLett.65.679",
    journal = "Phys. Rev. Lett.",
    volume = "65",
    pages = "679--682",
    year = "1990"
}

@article{Davidson:2000hf,
    author = "Davidson, Sacha and Hannestad, Steen and Raffelt, Georg",
    title = "{Updated bounds on millicharged particles}",
    eprint = "hep-ph/0001179",
    archivePrefix = "arXiv",
    reportNumber = "CERN-TH-99-384",
    doi = "10.1088/1126-6708/2000/05/003",
    journal = "JHEP",
    volume = "05",
    pages = "003",
    year = "2000"
}

@article{Mohapatra:1990vq,
    author = "Mohapatra, R. N. and Rothstein, I. Z.",
    title = "{ASTROPHYSICAL CONSTRAINTS ON MINICHARGED PARTICLES}",
    reportNumber = "UMDHEP-90-227",
    doi = "10.1016/0370-2693(90)91907-S",
    journal = "Phys. Lett. B",
    volume = "247",
    pages = "593--600",
    year = "1990"
}

@article{Vinyoles:2015khy,
    author = "Vinyoles, N{\'u}ria and Vogel, Hendrik",
    title = "{Minicharged Particles from the Sun: A Cutting-Edge Bound}",
    eprint = "1511.01122",
    archivePrefix = "arXiv",
    primaryClass = "hep-ph",
    reportNumber = "MPP-2015-199",
    doi = "10.1088/1475-7516/2016/03/002",
    journal = "JCAP",
    volume = "03",
    pages = "002",
    year = "2016"
}

@article{DeRocco:2020xdt,
    author = "DeRocco, William and Graham, Peter W. and Rajendran, Surjeet",
    title = "{Exploring the robustness of stellar cooling constraints on light particles}",
    eprint = "2006.15112",
    archivePrefix = "arXiv",
    primaryClass = "hep-ph",
    doi = "10.1103/PhysRevD.102.075015",
    journal = "Phys. Rev. D",
    volume = "102",
    number = "7",
    pages = "075015",
    year = "2020"
}

@article{CONNIE:2024off,
    author = "Aguilar-Arevalo, Alexis A. and others",
    collaboration = "CONNIE, Atucha-II",
    title = "{Search for Reactor-Produced Millicharged Particles with Skipper-CCDs at the CONNIE and Atucha-II Experiments}",
    eprint = "2405.16316",
    archivePrefix = "arXiv",
    primaryClass = "hep-ex",
    reportNumber = "FERMILAB-PUB-24-0709-PPD",
    doi = "10.1103/PhysRevLett.134.071801",
    journal = "Phys. Rev. Lett.",
    volume = "134",
    number = "7",
    pages = "071801",
    year = "2025"
}

@article{Ahlers:2007qf,
    author = "Ahlers, M. and Gies, H. and Jaeckel, J. and Redondo, J. and Ringwald, A.",
    title = "{Laser experiments explore the hidden sector}",
    eprint = "0711.4991",
    archivePrefix = "arXiv",
    primaryClass = "hep-ph",
    reportNumber = "DESY-07-207, OUTP-0715P, IPPP-07-93, DCPT-07-186",
    doi = "10.1103/PhysRevD.77.095001",
    journal = "Phys. Rev. D",
    volume = "77",
    pages = "095001",
    year = "2008"
}

@article{DellaValle:2014xoa,
    author = "Della Valle, F. and Milotti, E. and Ejlli, A. and Messineo, G. and Piemontese, L. and Zavattini, G. and Gastaldi, U. and Pengo, R. and Ruoso, G.",
    title = "{First results from the new PVLAS apparatus: A new limit on vacuum magnetic birefringence}",
    eprint = "1406.6518",
    archivePrefix = "arXiv",
    primaryClass = "quant-ph",
    doi = "10.1103/PhysRevD.90.092003",
    journal = "Phys. Rev. D",
    volume = "90",
    number = "9",
    pages = "092003",
    year = "2014"
}

@article{Lin:2023erb,
    author = "Lin, Rebecca and van Kerkwijk, Marten H. and Kirsten, Franz and Pen, Ue-Li and Deller, Adam T.",
    title = "{The Radio Parallax of the Crab Pulsar: A First VLBI Measurement Calibrated with Giant Pulses}",
    eprint = "2306.01617",
    archivePrefix = "arXiv",
    primaryClass = "astro-ph.HE",
    doi = "10.3847/1538-4357/acdc98",
    journal = "Astrophys. J.",
    volume = "952",
    number = "2",
    pages = "161",
    year = "2023"
}

@online{thompson_x-ray_nodate,
  title = {X-ray data booklet},
  author = "{A.~Thompson et~al.}",
  year = {2009},
  url = {https://xdb.lbl.gov/},
  urldate = {2019}
}

@article{Philippov:2013aha,
    author = "Philippov, Alexander and Tchekhovskoy, Alexander and Li, Jason G.",
    title = "{Time evolution of pulsar obliquity angle from 3D simulations of magnetospheres}",
    eprint = "1311.1513",
    archivePrefix = "arXiv",
    primaryClass = "astro-ph.HE",
    doi = "10.1093/mnras/stu591",
    journal = "Mon. Not. Roy. Astron. Soc.",
    volume = "441",
    number = "3",
    pages = "1879--1887",
    year = "2014"
}

@article{Lyne:1993xah,
    author = "Lyne, A. G. and Pritchard, R. S. and Graham Smith, F.",
    title = "{23 years of Crab pulsar rotational history}",
    doi = "10.1093/mnras/265.4.1003",
    journal = "Mon. Not. Roy. Astron. Soc.",
    volume = "265",
    number = "4",
    pages = "1003--1012",
    year = "1993"
}

@article{Manchester:2004bp,
    author = "Manchester, R N and Hobbs, G B and Teoh, A and Hobbs, M",
    title = "{The Australia Telescope National Facility pulsar catalogue}",
    eprint = "astro-ph/0412641",
    archivePrefix = "arXiv",
    reportNumber = "20041224",
    doi = "10.1086/428488",
    journal = "Astron. J.",
    volume = "129",
    pages = "1993",
    year = "2005"
}

@article{Hu:2016xas,
    author = "Hu, Ping-Kai and Kusenko, Alexander and Takhistov, Volodymyr",
    title = "{Dark Cosmic Rays}",
    eprint = "1611.04599",
    archivePrefix = "arXiv",
    primaryClass = "hep-ph",
    reportNumber = "IPMU16-0168, UCI-TR-2017-02",
    doi = "10.1016/j.physletb.2017.02.035",
    journal = "Phys. Lett. B",
    volume = "768",
    pages = "18--22",
    year = "2017"
}

@article{Dunsky:2018mqs,
    author = "Dunsky, David and Hall, Lawrence J. and Harigaya, Keisuke",
    title = "{CHAMP Cosmic Rays}",
    eprint = "1812.11116",
    archivePrefix = "arXiv",
    primaryClass = "astro-ph.HE",
    doi = "10.1088/1475-7516/2019/07/015",
    journal = "JCAP",
    volume = "07",
    pages = "015",
    year = "2019"
}

@article{Berlin:2018bsc,
    author = "Berlin, Asher and Blinov, Nikita and Krnjaic, Gordan and Schuster, Philip and Toro, Natalia",
    title = "{Dark Matter, Millicharges, Axion and Scalar Particles, Gauge Bosons, and Other New Physics with LDMX}",
    eprint = "1807.01730",
    archivePrefix = "arXiv",
    primaryClass = "hep-ph",
    reportNumber = "FERMILAB-PUB-18-310-A, SLAC-PUB-17297",
    doi = "10.1103/PhysRevD.99.075001",
    journal = "Phys. Rev. D",
    volume = "99",
    number = "7",
    pages = "075001",
    year = "2019"
}

@article{Berlin:2021kcm,
    author = "Berlin, Asher and Schutz, Katelin",
    title = "{Helioscope for gravitationally bound millicharged particles}",
    eprint = "2111.01796",
    archivePrefix = "arXiv",
    primaryClass = "hep-ph",
    reportNumber = "MIT-CTP/5358, FERMILAB-PUB-21-626-T",
    doi = "10.1103/PhysRevD.105.095012",
    journal = "Phys. Rev. D",
    volume = "105",
    number = "9",
    pages = "095012",
    year = "2022"
}

@article{Iles:2024zka,
    author = "Iles, Ella and Heeba, Saniya and Schutz, Katelin",
    title = "{Dark Matter Direct Detection Experiments Are Sensitive to the Millicharged Background}",
    eprint = "2407.21096",
    archivePrefix = "arXiv",
    primaryClass = "hep-ph",
    doi = "10.1103/PhysRevLett.134.121002",
    journal = "Phys. Rev. Lett.",
    volume = "134",
    number = "12",
    pages = "121002",
    year = "2025"
}

@article{Kelly:2018brz,
    author = "Kelly, Kevin J. and Tsai, Yu-Dai",
    title = "{Proton fixed-target scintillation experiment to search for millicharged dark matter}",
    eprint = "1812.03998",
    archivePrefix = "arXiv",
    primaryClass = "hep-ph",
    reportNumber = "FERMILAB-PUB-18-668-A-PPD-T",
    doi = "10.1103/PhysRevD.100.015043",
    journal = "Phys. Rev. D",
    volume = "100",
    number = "1",
    pages = "015043",
    year = "2019"
}

@article{Foroughi-Abari:2020qar,
    author = "Foroughi-Abari, Saeid and Kling, Felix and Tsai, Yu-Dai",
    title = "{Looking forward to millicharged dark sectors at the LHC}",
    eprint = "2010.07941",
    archivePrefix = "arXiv",
    primaryClass = "hep-ph",
    reportNumber = "FERMILAB-PUB-20-477-AE-PPD-T",
    doi = "10.1103/PhysRevD.104.035014",
    journal = "Phys. Rev. D",
    volume = "104",
    number = "3",
    pages = "035014",
    year = "2021"
}

@article{Tsai:2024wdh,
    author = "Tsai, Yu-Dai and Hwang, Insung and Schmitz, Ryan and Citron, Matthew and Gunthoti, Kranti and Steenis, Jacob and Jeong, Hoyong and Moon, Hyunki and Yoo, Jae Hyeok and Liu, Ming Xiong",
    title = "{LANSCE-mQ: Dedicated search for milli/fractionally charged particles at LANL}",
    eprint = "2407.07142",
    archivePrefix = "arXiv",
    primaryClass = "hep-ph",
    reportNumber = "FERMILAB-PUB-24-0357-T-V; LA-UR-24-27441, FERMILAB-PUB-24-0357-T-V",
    month = "7",
    year = "2024"
}

@article{Magill:2018tbb,
    author = "Magill, Gabriel and Plestid, Ryan and Pospelov, Maxim and Tsai, Yu-Dai",
    title = "{Millicharged particles in neutrino experiments}",
    eprint = "1806.03310",
    archivePrefix = "arXiv",
    primaryClass = "hep-ph",
    reportNumber = "FERMILAB-PUB-18-631-A",
    doi = "10.1103/PhysRevLett.122.071801",
    journal = "Phys. Rev. Lett.",
    volume = "122",
    number = "7",
    pages = "071801",
    year = "2019"
}

@article{Harnik:2020ugb,
    author = "Harnik, Roni and Plestid, Ryan and Pospelov, Maxim and Ramani, Harikrishnan",
    title = "{Millicharged cosmic rays and low recoil detectors}",
    eprint = "2010.11190",
    archivePrefix = "arXiv",
    primaryClass = "hep-ph",
    reportNumber = "FERMILAB-PUB-20-523-T",
    doi = "10.1103/PhysRevD.103.075029",
    journal = "Phys. Rev. D",
    volume = "103",
    number = "7",
    pages = "075029",
    year = "2021"
}

@article{Oscura:2023qch,
    author = "Perez, Santiago and others",
    collaboration = "Oscura",
    title = "{Searching for millicharged particles with 1 kg of Skipper-CCDs using the NuMI beam at Fermilab}",
    eprint = "2304.08625",
    archivePrefix = "arXiv",
    primaryClass = "hep-ex",
    reportNumber = "FERMILAB-PUB-23-200-PPD-T",
    doi = "10.1007/JHEP02(2024)072",
    journal = "JHEP",
    volume = "02",
    pages = "072",
    year = "2024"
}

@article{SENSEI:2023gie,
    author = "Barak, Liron and others",
    collaboration = "SENSEI",
    title = "{Search by the SENSEI Experiment for Millicharged Particles Produced in the NuMI Beam}",
    eprint = "2305.04964",
    archivePrefix = "arXiv",
    primaryClass = "hep-ex",
    reportNumber = "CALT-TH-2023-011, YITP-SB-2023-07, FERMILAB-PUB-23-222-PPD",
    doi = "10.1103/PhysRevLett.133.071801",
    journal = "Phys. Rev. Lett.",
    volume = "133",
    number = "7",
    pages = "071801",
    year = "2024"
}

@article{Plestid:2020kdm,
    author = "Plestid, Ryan and Takhistov, Volodymyr and Tsai, Yu-Dai and Bringmann, Torsten and Kusenko, Alexander and Pospelov, Maxim",
    title = "{New Constraints on Millicharged Particles from Cosmic-ray Production}",
    eprint = "2002.11732",
    archivePrefix = "arXiv",
    primaryClass = "hep-ph",
    reportNumber = "FERMILAB-PUB-20-044-A-T, INT-PUB-20-004, IPMU20-0015",
    doi = "10.1103/PhysRevD.102.115032",
    journal = "Phys. Rev. D",
    volume = "102",
    pages = "115032",
    year = "2020"
}

@article{Cohen:2008wz,
    author = "Cohen, Thomas D. and McGady, David A.",
    title = "{The Schwinger mechanism revisited}",
    eprint = "0807.1117",
    archivePrefix = "arXiv",
    primaryClass = "hep-ph",
    doi = "10.1103/PhysRevD.78.036008",
    journal = "Phys. Rev. D",
    volume = "78",
    pages = "036008",
    year = "2008"
}

@article{Nikishov:1970br,
    author = "Nikishov, A. I.",
    title = "{Barrier scattering in field theory removal of klein paradox}",
    doi = "10.1016/0550-3213(70)90527-4",
    journal = "Nucl. Phys. B",
    volume = "21",
    pages = "346--358",
    year = "1970"
}

@article{Du:2022hms,
    author = "Du, Mingxuan and Fang, Rundong and Liu, Zuowei",
    title = "{Millicharged particles from proton bremsstrahlung in the atmosphere}",
    eprint = "2211.11469",
    archivePrefix = "arXiv",
    primaryClass = "hep-ph",
    doi = "10.1007/JHEP08(2024)174",
    journal = "JHEP",
    volume = "08",
    pages = "174",
    year = "2024"
}

@article{Caputo:2019tms,
    author = "Caputo, Andrea and Sberna, Laura and Frias, Miguel and Blas, Diego and Pani, Paolo and Shao, Lijing and Yan, Wenming",
    title = "{Constraints on millicharged dark matter and axionlike particles from timing of radio waves}",
    eprint = "1902.02695",
    archivePrefix = "arXiv",
    primaryClass = "astro-ph.CO",
    doi = "10.1103/PhysRevD.100.063515",
    journal = "Phys. Rev. D",
    volume = "100",
    number = "6",
    pages = "063515",
    year = "2019"
}

@article{XENON:2022ltv,
    author = "Aprile, E. and others",
    collaboration = "XENON",
    title = "{Search for New Physics in Electronic Recoil Data from XENONnT}",
    eprint = "2207.11330",
    archivePrefix = "arXiv",
    primaryClass = "hep-ex",
    doi = "10.1103/PhysRevLett.129.161805",
    journal = "Phys. Rev. Lett.",
    volume = "129",
    number = "16",
    pages = "161805",
    year = "2022"
}

@article{Berlin:2020pey,
    author = "Berlin, Asher and Hook, Anson",
    title = "{Searching for Millicharged Particles with Superconducting Radio-Frequency Cavities}",
    eprint = "2001.02679",
    archivePrefix = "arXiv",
    primaryClass = "hep-ph",
    doi = "10.1103/PhysRevD.102.035010",
    journal = "Phys. Rev. D",
    volume = "102",
    number = "3",
    pages = "035010",
    year = "2020"
}

@article{Caputo:2021eaa,
    author = "Caputo, Andrea and Millar, Alexander J. and O'Hare, Ciaran A. J. and Vitagliano, Edoardo",
    title = "{Dark photon limits: A handbook}",
    eprint = "2105.04565",
    archivePrefix = "arXiv",
    primaryClass = "hep-ph",
    reportNumber = "NORDITA-2021-036",
    doi = "10.1103/PhysRevD.104.095029",
    journal = "Phys. Rev. D",
    volume = "104",
    number = "9",
    pages = "095029",
    year = "2021"
}

@article{Ruderman:1975ju,
    author = "Ruderman, M. A. and Sutherland, P. G.",
    title = "{Theory of pulsars: Polar caps, sparks, and coherent microwave radiation}",
    doi = "10.1086/153393",
    journal = "Astrophys. J.",
    volume = "196",
    pages = "51",
    year = "1975"
}

@article{Kim:2003qp,
    author = "Kim, Sang Pyo and Page, Don N.",
    title = "{Schwinger pair production in electric and magnetic fields}",
    eprint = "hep-th/0301132",
    archivePrefix = "arXiv",
    doi = "10.1103/PhysRevD.73.065020",
    journal = "Phys. Rev. D",
    volume = "73",
    pages = "065020",
    year = "2006"
}

@article{ArguellesDelgado:2021lek,
    author = {Arg{\"u}elles Delgado, Carlos Alberto and Kelly, Kevin James and Mu{\~n}oz Albornoz, V{\'\i}ctor},
    title = "{Millicharged particles from the heavens: single- and multiple-scattering signatures}",
    eprint = "2104.13924",
    archivePrefix = "arXiv",
    primaryClass = "hep-ph",
    reportNumber = "FERMILAB-PUB-21-214-T",
    doi = "10.1007/JHEP11(2021)099",
    journal = "JHEP",
    volume = "11",
    pages = "099",
    year = "2021"
}

@article{Gao:2025ykc,
    author = "Gao, Ting and Pospelov, Maxim",
    title = "{Constraints on millicharged particles from nuclear gamma-decays}",
    eprint = "2507.17955",
    archivePrefix = "arXiv",
    primaryClass = "hep-ph",
    month = "7",
    year = "2025"
}

@article{Wu:2024iqm,
    author = "Wu, Han and Hardy, Edward and Song, Ningqiang",
    title = "{Searching for heavy millicharged particles from the atmosphere}",
    eprint = "2406.01668",
    archivePrefix = "arXiv",
    primaryClass = "hep-ph",
    doi = "10.1103/PhysRevD.110.115037",
    journal = "Phys. Rev. D",
    volume = "110",
    number = "11",
    pages = "115037",
    year = "2024"
}

@article{PhysRevD.49.2114,
  title = {Astrophysical bounds on millicharged particles in models with a paraphoton},
  author = {Davidson, Sacha and Peskin, Michael},
  journal = {Phys. Rev. D},
  volume = {49},
  issue = {4},
  pages = {2114--2117},
  numpages = {0},
  year = {1994},
  month = {Feb},
  publisher = {American Physical Society},
  doi = {10.1103/PhysRevD.49.2114},
  url = {https://link.aps.org/doi/10.1103/PhysRevD.49.2114}
}

@article{Li:2024pcp,
    author = "Li, Changqian and Liu, Zuowei and Lu, Wenxi and Ye, Zicheng",
    title = "{Low-energy supernova constraints on millicharged particles}",
    eprint = "2408.04953",
    archivePrefix = "arXiv",
    primaryClass = "hep-ph",
    doi = "10.1007/JHEP07(2025)116",
    journal = "JHEP",
    volume = "07",
    pages = "116",
    year = "2025"
}

@article{Chang:2018rso,
    author = "Chang, Jae Hyeok and Essig, Rouven and McDermott, Samuel D.",
    title = "{Supernova 1987A Constraints on Sub-GeV Dark Sectors, Millicharged Particles, the QCD Axion, and an Axion-like Particle}",
    eprint = "1803.00993",
    archivePrefix = "arXiv",
    primaryClass = "hep-ph",
    reportNumber = "YITP-SB-18-01, FERMILAB-PUB-17-432-T",
    doi = "10.1007/JHEP09(2018)051",
    journal = "JHEP",
    volume = "09",
    pages = "051",
    year = "2018"
}

@article{Gan:2023jbs,
    author = "Gan, Xucheng and Tsai, Yu-Dai",
    title = "{Cosmic millicharge background and reheating probes}",
    eprint = "2308.07951",
    archivePrefix = "arXiv",
    primaryClass = "hep-ph",
    reportNumber = "UCI-HEP-TR-2023-05, FERMILAB-PUB-23-428-T-V",
    doi = "10.1007/JHEP07(2025)094",
    journal = "JHEP",
    volume = "07",
    pages = "094",
    year = "2025"
}

@article{Holdom:1985ag,
    author = "Holdom, Bob",
    title = "{Two U(1)'s and Epsilon Charge Shifts}",
    reportNumber = "UTPT-85-30",
    doi = "10.1016/0370-2693(86)91377-8",
    journal = "Phys. Lett. B",
    volume = "166",
    pages = "196--198",
    year = "1986"
}

\end{document}